\documentstyle[fleqn,espcrc2,epsf]{article}
\title{$B \rightarrow \pi l\overline{\nu}$ Form Factors
with NRQCD Heavy Quark and Clover Light Quark Actions
\thanks{presented by T. Onogi}} 

\author{JLQCD Collaboration:
        S.~Aoki\address{Institute of Physics, University of Tsukuba,
        Tsukuba, Ibaraki 305-8571, Japan},
        M.~Fukugita\address{Institute for Cosmic Ray Research, University
        of Tokyo, Tanashi, Tokyo 188-8502, Japan},
        S.~Hashimoto\address{High Energy Accelerator Research Organization
        (KEK), Tsukuba, Ibaraki305-0801, Japan},
        K-I.~Ishikawa$^{\rm c}$,
        N.~Ishizuka$^{\rm a,}$\address{Center for Computational Physics,
        University of Tsukuba, Tsukuba, Ibaraki 305-8577, Japan},
        Y.~Iwasaki$^{\rm a,d}$,
        K.~Kanaya$^{\rm a,d}$,
        T.~Kaneda$^{\rm a}$,
        S.~Kaya$^{\rm c}$,
        Y.~Kuramashi$^{\rm c}$,
        M.~Okawa$^{\rm c}$,
        T.~Onogi\address{Department of Physics, Hiroshima University,
        Higashi-Hiroshima, Hiroshima 739-8526, Japan},
        S.~Tominaga$^{\rm c}$,
        N.~Tsutsui$^{\rm e}$,
        A.~Ukawa$^{\rm a,d}$,
        N.~Yamada$^{\rm e}$,
        and
        T.~Yoshi\'e$^{\rm a,d}$}

\begin{document}
\begin{abstract}
  We report results on semileptonic $B\rightarrow\pi l\overline{\nu}$
  decay form factors near $q^2_{\rm max}$ using NRQCD heavy quark
  and clover light quark actions and currents improved through 
  $O(\alpha a)$. 
  An inconsistency with the soft pion relation 
  $f^0(q^2_{\rm max})=f_B/f_{\pi}$ found in a previous work is confirmed, and a 
  possible solution with nonperturbative renormalization is
  discussed. 
  We find that $f^+(q^2)$ is well described by the $B^*$ pole near
  $q^2_{\rm max}$, and its $1/M_B$ scaling is also consistent
  with the prediction of the pole dominance model.
\end{abstract}
\maketitle

\section{Introduction}
The $B\rightarrow\pi l\overline{\nu}$ form factors are relevant for the 
extraction of the CKM matrix element $|V_{ub}|$ through the
exclusive decay.
While lattice QCD computation can cover only the region
near $q^2=q^2_{\rm max}$ with reasonable statistical and
discretization errors, it is still useful once the experiments
reach sufficiently high statistics to measure the partial decay
rate in the same region.
In this article we report preliminary results of our study of the 
$B\rightarrow\pi l\overline{\nu}$ form factors using NRQCD for heavy quark.

\section{Simulation}
We employ $O(1/M)$ NRQCD for heavy quark.  For the light quark 
the SW clover quark action is used, with 
the clover coefficient $c_{\rm sw}$ determined by mean
field improved perturbation theory at one-loop order. 
The heavy-light vector current involved in the matrix
element is renormalized to $O(\alpha_s a)$ 
using the one-loop calculation of
Morningstar and Shigemitsu \cite{MS99}, and of Ishikawa
\cite{I99} including mixings with higher dimensional
operators. 

The calculation of $f^0(q_{\rm max})$ was performed at 
$\beta$=5.7 and 5.9 on $12^3 \times 32$ and  
$16^3 \times 40$ lattices  with 232 and 222
gauge configurations, respectively.
We took five different heavy quark masses covering $m_b$ and
four different light quark masses ranging from $m_s$ to
$m_s/2$. 
For $f^+(q^2)$ we only analyzed the $\beta=5.7$ data, so
far.
The momentum combinations $p_B$=$(0,0,0)$,
$k_{\pi}$=$(0,0,0)$, $(1,0,0)$, $(1,1,0)$, and
$p_B$=$(1,0,0)$, $k_{\pi}$=$(0,0,0)$ in units of $2\pi/(12 a)$ 
are considered
at $\kappa_l=0.1369$, which is around $m_s$.

\section{Results for $f^0(q^2_{\rm max})$}
\begin{figure}[t]
  \vspace*{-0.6cm}
  \centerline{
    \epsfxsize=7.0cm 
    \epsfbox{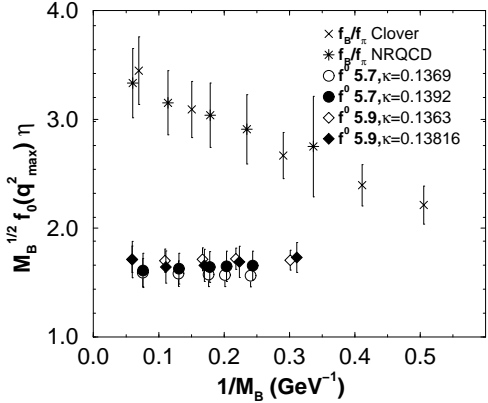}
    }
  \vspace*{-9mm}
  \caption{ Comparison of $\sqrt{M_B} f^0(q^2_{\max}) \eta $
    with $\sqrt{M_B} f_B/f_{\pi} \eta $, where $\eta$ in defined as
    $(\alpha_s(M_B)$ 
    $/\alpha_s(M_B^{\rm phys}))^{2/\beta_0}$. 
    $f^0(q^2_{\rm max})$ with heaviest and with lightest
    light quark mass is plotted at $\beta$=5.7 and 5.9.
    Data for $f_B/f_{\pi}$ are taken from our previous
    works: heavy clover \cite{JLQCD_fB_clover} and NRQCD
    \cite{JLQCD_fB_NRQCD}. 
    }
  \label{softpion}
  \vspace*{-0.6cm}
\end{figure}

Let us first present the results for $f^0(q^2_{\rm max})$,
for which the soft pion theorem 
$f^0(q^2_{\rm max})=f_B/f_{\pi}$ in the chiral limit
provides an important check of the lattice calculation.
The $1/M_B$ dependence of 
$\sqrt{M_B}$ $f^0(q^2_{\rm max})$
 $(\alpha_s(M_B)$ $/\alpha_s(M_B^{\rm phys}))^{2/\beta_0}$ 
and a comparison with
$\sqrt{M_B}$ $f_B/f_{\pi}$ $(\alpha_s(M_B)$ 
$/\alpha_s(M_B^{\rm phys}))^{2/\beta_0}$
is shown in Fig.~\ref{softpion}.
The data for $f^0(q^2_{\rm max})$ at two $\beta$ values show 
nice scaling, while a clear disagreement with $f_B/f_{\pi}$ is
observed, confirming the point made by the Hiroshima Group
\cite{Matsufuru}. 

The chiral (soft pion) limit is not taken for the points
given in the plot, which is a possible reason of the
violation of the soft pion relation.
We find, however, that the light quark mass dependence of
$f^0(q^2_{\rm max})$ is consistent with a constant within
statistical error, and a polynomial chiral extrapolation in 
$m_q$ and $m_q^2$ gives a consistent result.
With the present statistics, we are not able to fit our data with 
both $\sqrt{m_q}$ and $m_q$ in the fitting function.

Another possible reason for the disagreement with the soft
pion relation is the large uncertainty in the matching
constants. 
Since the vector and axial vector heavy-light currents are involved 
on the two sides of the equality, perturbative errors in the
matching between the continuum and lattice operators could be
important.
Naively this error is $O(\alpha_s^2)$,
and hence should be small at $\beta\sim$ 6.0. 
Nonetheless, the large one-loop
correction in the renormalization constant $Z_A^{HL}$
suggests that there could be large higher order
corrections. 

In order to see how such higher order effect contributes, we
computed the ratio of the renormalization constants
$Z_A^{HL}/Z_V^{HL}$ nonperturbatively in the static limit,
using the chiral Ward-Takahashi identity 
\begin{eqnarray}
  \label{eq:chiral_Ward_identity}
  \lefteqn{
    Z_A Z_V^{HL} \int d^4y \langle
    (\partial_{\mu}A_{\mu}-2m_q P)(y) V_0^{HL}(x)
    \cal{O} \rangle 
    } \nonumber \\
    & = & - Z_A^{HL} \langle A_0^{HL}(x) \cal{O} \rangle,
\end{eqnarray}
where $A_{\mu}$ and $P$ denote light-light axial-current and 
pseudoscalar density, with $Z_A$
the renormalization factor for $A_{\mu}$, and 
 $V_0^{HL}$ and $A_0^{HL}$
are the heavy-light (static-light in this particular case)
currents.
We performed simulations on a $12^3 \times 32$ lattice at $\beta=6.0$
following the methods of Maiani-Martinelli
\cite{Maiani_Martinelli_86}. For the operator $\cal{O}$ we took
a heavy-light meson interpolation operator with wall source.
The clover coefficient $c_{sw}$ for the light quark was chosen 
to be the nonperturbative value from \cite{ALPHA_96}

Combining our result for $Z_A^{HL}/(Z_V^{HL} Z_A)$ from 
(1) with the nonperturbative value of \cite{ALPHA_96} for $Z_A$, 
we obtained $Z_A^{HL}/Z_V^{HL}=0.72(2)$ to be compared with 
the perturbative result $0.87(3)$. 
We find that the nonperturbative value for 
$Z_A^{HL}/Z_V^{HL}$ is about 20\% smaller than the
corresponding one-loop result.  While this explains part of the
discrepancy between $f_B/f_{\pi}$ and $f^0(q^2_{\rm max})$, 
the reduction is not sufficient to remove the disagreement
seen in Fig.~\ref{softpion}. 

In our study of the renormalization constant, the light-light 
and static-light current did not include corrections from 
higher dimensional operators.  Since these corrections  
are known to give a large contribution in the calculation of 
$f_B$ \cite{MS99}, it would be important to perform 
a study of renormalization constant with improved curents.

\section{Results for $f^+(q^2)$}

\begin{figure}[t]
  \vspace*{-0.6cm}
  \centerline{\epsfxsize=7.0cm 
    \epsfbox{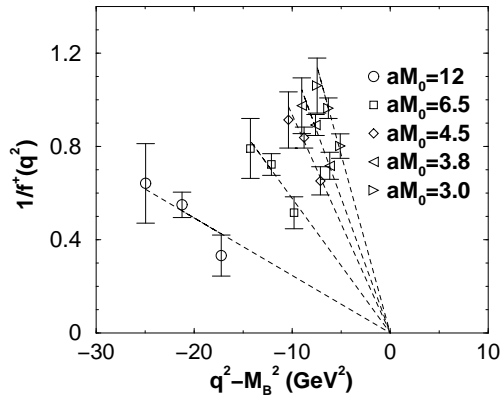}}
  \vspace*{-9mm}
  \caption{
    Pole fit of $1/f^+(q^2)$.
    Lattice results at five different heavy quark mass are
    shown at $\beta$=5.7.
    The $B^*$ pole is almost degenerated with $B$, and
    located at the origin in the plot. $a M_0$ is the 
    bare heavy quark mass.
    }
  \label{pole}
  \vspace*{-0.6cm}
\end{figure}

We next study the $q^2$ and $1/M_B$ dependence of the form
factor $f^+(q^2)$. 
For the large $q^2$ region, the form factor $f^+(q^2)$
should be well approximated by the pole dominance model with 
the $B^*$ pole, since $B^*$ is almost degenerate with $B$ in
the heavy quark limit and the pole is very close to
$q^2_{\rm max}$.
The pole model predicts
\begin{equation}
  \label{eq:pole_model}
  1/f^+(q^2) =  - c_1 (q^2 - M_{B^*}^2)
  + O((q^2 - M_{B^*}^2)^2),
\end{equation}
where the coefficient $c_1$ is written in terms of the
dimensionless $B^*B\pi$ coupling $g$ and the $B^*$ meson
decay constant $f_{B^*}$ as
$c_1 = f_{\pi}/(f_{B^*} M_{B^*}^2 g)$.
Figure~\ref{pole} shows our data for $1/f^+(q^2)$ near
$q^2_{\rm max}$ for $\kappa=0.1369$ at $\beta=5.7$.
We find that the pole fit indeed explains the data quite
well for each value of the heavy quark mass.

\begin{figure}[t]
  \vspace*{-0.6cm}
  \centerline{\epsfxsize=7.0cm 
    \epsfbox{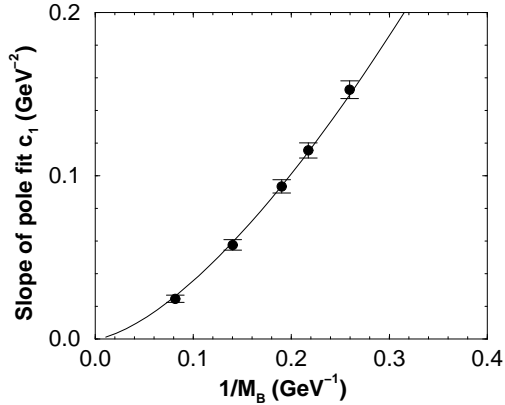}}
  \vspace*{-9mm}
  \caption{ $1/M_B$ scaling of the pole fit coefficient.}
  \label{cfp}
  \vspace*{-0.2cm}
\end{figure}

The heavy quark scaling is also predicted within the pole
dominance model.
Using $f_{B^*}\sim M_B^{-1/2}$ and $g\sim$ constant, we obtain
$c_1\sim M_B^{-3/2}$.
The slope obtained with the fit (\ref{eq:pole_model}) is
plotted against $1/M_B$ in Fig.~\ref{cfp} together with a
curve representing $1/M_B^{3/2}$.
We confirm that the heavy quark scaling is nicely satisfied.
Furthermore, from this fit, we obtain $g$=0.33(4), which is
consistent with the value extracted from 
$D^*\rightarrow D\pi$ \cite{DDpi} $g=0.27(6)$ and with the recent
lattice study \cite{UKQCD} $g=0.42(8)$.

\section{Conclusions}
We consider that the problem of the soft pion relation should be 
resolved before a prediction of the $B\rightarrow\pi l\overline{\nu}$ form factors
from lattice QCD can be made.  
We have investigated the possibility that the perturbative matching 
contains a large systematic error, and found that a nonperturbative 
value shows a non-negligible difference from the one-loop result.

Checking the $q^2$ dependence and the $1/M_B$ scaling of
$f^+(q^2)$ is a necessary step toward a prediction of
partial decay rate.  We have found that the pole
dominance model provides an excellent way to fit
the observed shape of $f^+(q^2)$ in the large $q^2$ region.
The $1/M$ scaling is also consistent with the pole model
prediction. 

\vspace{6mm}
This work is supported by the Supercomputer Project No.45
(FY1999) of High Energy Accelerator Research Organization
(KEK), and also in part by the Grants-in-Aid of the Ministry
of Education (Nos. 09304029, 10640246, 10640248, 10740107,
10740125, 11640294, 11740162).  
K-I.I is supported by the JSPS Research Fellowship. 

\end{document}